\documentclass[a4paper,fleqn,usenatbib]{mnras}

\usepackage{newtxtext,newtxmath}

\usepackage[T1]{fontenc}
\usepackage{ae,aecompl}

\usepackage{graphicx}	
\usepackage{amsmath}	
\usepackage{amssymb}	

\title[The quench of galaxies in A\ 85 and its vicinity]{Deep spectroscopy of nearby galaxy clusters: IV The quench of the star formation in galaxies in the infall region of Abell\ 85}

\author[J. A. L. Aguerri et al.]{J. A. L. Aguerri$^{1,2}$\thanks{jalfonso@iac.es}, I. Agulli$^{1,2,3,4}$\thanks{ireagu@iac.es}, and J. M\'endez-Abreu$^{1,2}$  \\
$^{1}$ Instituto de Astrof\'{\i}sica de Canarias, C/ V\'{\i}a L\'actea s/n, E-38205, La Laguna, Spain\\
$^{2}$ Departamento de Astrof\'{\i}sica, Universidad de La Laguna, E-38206, La Laguna, Spain\\
$^{3}$ Dipartimento di Fisica, Universit\`a di Torino, Via P. Giuria 1, I-10125 Torino, Italy \\
$^{4}$ Istituto Nazionale di Fisica Nucleare (INFN), sezione di Torino, Via P. Giuria 1, I-10125 Torino, Italy
}

\date{Accepted XXX. Received YYY; in original form ZZZ}

\pubyear{2016}

\begin{document}
\label{firstpage}
\pagerange{\pageref{firstpage}--\pageref{lastpage}}
\maketitle

\begin{abstract}
Our aim is to understand the role of the environment in the quenching of star formation of galaxies located in the infall cluster region of Abell 85 (A\ 85). This is achieved by studying the post-starburst galaxy population as tracer of recent quenching. By measuring the equivalent width (EW) of the [OII] and H$\delta$ spectral lines, we classify the galaxies in three groups: passive (PAS), emission line (EL), and post-starburst (PSB) galaxies.  The PSB galaxy population represents $\sim 4.5\%$ of the full sample.  Dwarf galaxies ($M_{r} > -18.0$) account for $\sim 70 - 80\%$ of PSBs, which indicates that most of the galaxies undergoing recent quenching are low-mass objects. 
Independently of the environment, PSB galaxies are disk-like objects with  $g - r$  colour between the blue ELs and the red PAS ones. The PSB and EL galaxies in low-density environments show similar luminosities and local galaxy densities. The dynamics and local galaxy density of the PSB population in high density environments are shared with PAS galaxies. However, PSB galaxies inside A\ 85 are at shorter clustercentric radius than PAS and EL ones. The value of the EW(H$\delta$) is larger for those PSBs closer to the cluster centre. We propose  two different physical mechanisms producing PSB galaxies depending on the environment. In low density environments,  gas-rich minor mergers or accretions could produce the PSB galaxies.  For high density environments like A\ 85, PSBs  would be produced by the removal of the gas reservoirs of EL galaxies by ram-pressure stripping when they pass near to the cluster centre.
\end{abstract}



\section{Introduction}

One of the main results obtained during the last two decades by several large scale galaxy surveys  is the stellar colour dichotomy observed  in the colour-magnitude or colour-mass diagrams (CMD) for the low-redshift galaxy population. According to their stellar colour, galaxies can be grouped in either  blue or red ones \citep[e.g.,][]{strateva2001,hogg2004,baldry2006}.   Red galaxies are located in the so-called red sequence of the CMDs. They are objects with few or no gas content, non-active star formation, and with stellar populations evolving passively. This red sequence galaxies have mainly early-type morphologies and are located in high galaxy density environments \citep[e.g.,][]{hogg2004, balogh2004,sanchezalmeida2011,aguerri2012}. Blue galaxies are forming  the blue cloud of the CMD. They are gas-rich objects with active star formation, mainly showing late-type morphologies, and primary located in low density environments \citep[e.g.,][]{blanton2003,hogg2004,balogh2004, brichmann2004}.
 
The colour bimodality is kept up to high redshift ($z\sim1$), although there is a large evolution in the build up of the red sequence. Thus, the red sequence is in place for bright galaxies at $z \sim$ 1, while at the dwarf regime ($M_{r} > -18.0$)  the build up happened at smaller redshifts \citep[e.g.,][]{bundy2005, delucia2007}. Indeed, some low redshift environments are still building up the red sequence in the dwarf regime \citep[][]{agulli2016a}. The evolution with redshift of the red sequence shows that galaxies have evolved from the blue cloud to the red sequence by quenching their star formation and reddening their stellar colours \citep[e.g.,][]{bell2004,faber2007}. The quench of the star formation in galaxies can be driven by different mechanisms. In general they can be grouped in internal processes (self quench) or external ones \citep[environmental quench; e.g.,][]{peng2010}. The stellar mass is thought to be crucial for the dominance of internal or external quenching mechanisms in galaxies \citep[][]{peng2010}. Understanding these two quenching modes is one of the key problems in the modern extragalactic astronomy.

The role played by high-density environments in the quenching of galaxies is supported by several observational evidences. The most important are the morphology - density relation \citep[e.g.,][]{dressler1980,dressler1997,fasano2015}, the Butcher - Oemler effect \citep[e.g.,][]{butcher1984, aguerri2007}, or the colour - density relation \citep[e.g.,][]{hogg2004, balogh2004, sanchezjanssen2008}. These relations are linked with other trends observed in cluster member galaxies related with other galaxy properties. We can mention the dependence with the environment of the structures of galaxies \citep[e.g.,][]{trujillo2001, aguerri2004, gutierrez2004, mendezabreu2012, aguerri2016}, or the star formation rate \citep[e.g.,][]{lewis2002, haines2015}.

The different properties of the galaxies observed both in clusters and  in the field are produced by a number of physical mechanisms acting in high density environments. The large number of galaxies in clusters and their large relative velocity produce high speed encounters between galaxies and with the cluster gravitational potential. This is the so-called galaxy harassment that has been proposed to induce strong morphological changes in galaxies \citep[][]{moore1998, mastropietro2005, aguerri2009}. In the centre of the clusters interactions between galaxies and with the cluster potential could  produce mergers \citep[][]{mihos1994} or even destroy the galaxies producing the so-called intracluster light observed in some nearby galaxy clusters \citep[e.g.,][]{arnaboldi2002, aguerri2005, mihos2005, castrorodriguez2009}. These gravitational mechanisms affect  both the stellar and the gas content of the galaxies in clusters. Other mechanisms only affect their gas reservoirs. The ram-pressure stripping,  produced by the interaction of the galaxies with the hot intracluster medium,  sweeps off their cold gas content \citep[e.g.,][]{gunn1972, quilis2000}. In addition, strangulation \citep[e.g.,][]{larson1980,peng2015} and starvation \citep[][]{balogh2000,vandenvoort2017} produce the loss of the hot gas located in their halos. The result of all these hydrodynamical mechanisms is the loss of the galaxy's gas reservoir and the subsequent quench of their star formation.

All galaxies located in clusters evolve under the previous physical mechanisms. Nevertheless, the effects produced by those mechanisms depend on many parameters like galaxy mass, orbits, or morphology \citep[e.g.,][]{smith2015}. In addition, their different time-scales and zones of influence in the clusters make difficult to infer which is the main mechanism producing the quench of the star formation, and therefore how is the  evolution of the galaxies in clusters \citep[e.g.,][]{treu2003}. The study of the properties of galaxies that have recently stopped their star formation can help us to understand how the star forming galaxy population turns passive. These recently quenched objects exist and are called post-starburts (PSB) galaxies. 

\cite{dressler1982} found a number of galaxies located in galaxy clusters with peculiar features in their spectra. In particular, they show no emission lines and strong Balmer absorption lines. The absence of emission lines in their spectra indicates that the massive O- and B-type stars populations  have die. Nevertheless, the strong Balmer absortion lines show that A-type stars are still present. These features are typical of stellar populations that have abruptly stopped their star formation within the last 1 - 1.5 Gyr \citep[e.g.,][]{couch1987, abraham1996, poggianti1997, poggianti2004}. The spectra of these galaxies can be decomposed by using a combination of a K-type giant star and an A-type star spectrum \citep[e.g.,][]{quintero2004}. This is the reason why these objects are also called $k + a$ or $a + k$ galaxies in the literature.

Several surveys show that a  fraction of PSB galaxies are located in the field rather than in clusters \citep{zabludoff1996, goto2003, quintero2004, balogh2005, cybulski2014}. This has been interpreted as the main mechanism producing PSB galaxies is not related with the cluster environment. Galaxy-galaxy interactions or mergers of gas-rich galaxies in poor galaxy groups could be the responsible of the formation of the PSB spectral features in low density environments \citep[e.g.,][]{barnes1991, bekki2001, bekki2005, wild2009, snyder2011}. Other surveys did not find such difference between  clusters and field. \cite{balogh1999} concluded that the cluster environment inhibit the activation of starbursts rather than the quench of the star formation. The debate on the connection between the PSB galaxies and galaxy-galaxy interactions or mergers is still open \citep[e.g.,][]{pawlik2016}. In addition, the quenching mechanisms could vary with redshift \citep[e.g.,][]{wild2016}.

PSB galaxies were first observed by \cite{dressler1983} in a medium redshift cluster survey. Their population seems to be mainly in clusters at high redshift \citep[e.g.,][]{dressler1999, poggianti2009}. \cite{muzzin2014} found that the phase-space distribution of the PSB galaxies in a high redshift cluster indicates that they were quenched in a short time-scale ($\tau \sim 0.5\ Gyr$) after the first cross 0.5$R_{200}$ of the cluster. In nearby clusters, PSB galaxies represents up to $\sim 7\%$ of the galaxy cluster population within $\sim R_{200}$. Their fraction increases from the outskirts (less dense regions) to the cluster centre (high dense regions). These fractions also depend on other properties of the cluster. Their position in the phase-space indicates that PSB galaxies are objects with different accretion histories \citep[][]{paccagnella2017}. \cite{poggianti2004} observed the conection between PSB galaxies in the Coma cluster with the position of strong X-ray temperature gradients. The high velocity dispersion in galaxy clusters makes  mergers less efficient for the formation of PSB galaxies in clusters. In contrast, the central location of PSBs and their short quenching time-scales suggest that other mechanisms like ram-pressure stripping could produce them \citep[][]{muzzin2014, paccagnella2017}.  \cite{dressler2013} provided a global picture of the PSB phenomenom in different environments.  They suggested that starburst and post-starburst galaxies are indications of minor mergers and accretions in star forming and passive populations, respectively.

In the present paper we have analyzed the properties of the PSB galaxy population in both the cluster and the infall region of  Abell\ 85 (A\ 85). Several reasons make this nearby ($z = 0.055$) and massive \citep[$M_{\odot}= 2.5 \times 10^{14} M_{\odot} $;][]{rines2006} galaxy cluster  an ideal framework  to test galaxy transformations with the environment. First, several spectroscopic studies have been focused on this cluster \citep[e.g., ][]{durret1998a, rines2006, aguerri2007, aguerri2010, agulli2014, agulli2016a}.  They provide us with a large number of cluster members (460 galaxies) and galaxies in the infall cluster region ($\sim 700$ galaxies with  V $< 32000$ km s$^{-1}$) down to $\sim M^{*} + 6$ \citep[][]{agulli2016b}. Second, this cluster shows a large diversity of environments. In particular, along the line-of-sight of this cluster can be observed a filamentary galaxy structure connecting A\ 85 with other smaller and nearby groups \citep[][]{durret1998b}. The richness of environments and the deep spectroscopy available allow us to analyze the galaxy evolution at different galaxy densities down to the dwarf regime. 

The galaxy luminosity function (LF) of A85 also presents some features that makes  this environment special. In particular, the LF of this cluster shows an upturn at its faint-end. Nowadays, this is the only upturn spectroscopically confirmed in the LF of nearby clusters \citep[][]{agulli2014}. However, this upturn is not as steep as the observed in photometric galaxy LFs in clusters \citep[see ][]{popesso2006}. In addition, the faint-end slope of the LF in A85 is similar to the field one. The faint-end of the A85 LF is dominated in number by red dwarf galaxies. In contrast, blue dwarf systems dominate the faint-end of the field LF \citep[][]{agulli2014, agulli2016b}. These results suggest that an important quenching of the star formation happens for blue dwarf galaxies from the field to the cluster environment. But, is this quenching produced inside the cluster enviroment?;  are blue dwarf galaxies transformed in the infall cluster region?;  which are the mechanisms driving the quenching of the galaxies in the A\ 85 environment?. To answer these questions we have studied the properties of the PSB galaxies in the infall cluster region of A\ 85.
  
 The paper is organized as follows. Section 2 shows the data used, the large scale structure around the cluster, the procedure to compute the local galaxy density, and the galaxy spectral classification. Section 3  describes our main results. The discussion and conclusions are given in Sects. 4 and 5, respectively. Through this paper we have used the cosmology given by $\Omega_{\Lambda} = 0.7$, $\Omega_{m}=0.3$, and $H_{0}=75$ Mpc$^{-1}$ km s$^{-1}$.


\section{Spectroscopic data and large scale structure of Abell\ 85}

\subsection{Spectroscopic data from A\ 85}

The galaxy cluster A\ 85 has been observed in several optical spectroscopic surveys in the literature \citep[][]{durret1998b,aguerri2007,aguerri2010,boue2008,bravoalfaro2009}. In addition, the sky region around this cluster has been surveyed by the Sloan Digital Sky Survey \citep[SDSS; ][]{eisenstein2011} providing photometry in the $u, g, r, i$, and $z$  filters and spectroscopy of the galaxies down to $m_{r} \sim 18.0$. 

We have completed these studies by running a deeper spectroscopic survey focussed on galaxies with  $m_{r} > 18.0$, and  not previously observed \citep[][]{agulli2014,agulli2016b}. Our observations covers an area of $3.0 \times 2.6$  Mpc$^{2}$ around the cluster centre. This corresponds to a radius of 1.4 $r_{200}$ \citep[][]{agulli2014, agulli2016b}. The spectroscopy was obtained in two campaings by using the VIMOS@VLT and the AF2@WHT spectrographs. The reduction and calibration of the data were done by using the standard data reduction pipelines of the instruments \citep[][]{izzo2004,dominguezpalmero2014}. We refer the reader to  \cite{agulli2016b} for more details. 

The combination of  data from both the literature and our observations result in a total of 1603 redshifts in the magnitude interval $13 \leq m_{r} \leq 22$, within 1.4 $r_{200}$, and for galaxies with $g - r < 1.0$. This is the typical colour of a 12 Gyr old stellar population with $[Fe/H] =+0.25$ supersolar metallicity  \citep[][]{worthey1994}, representative of very luminous early-type galaxies. As a result, this colour selection should minimize the contamination due to background sources, while at the same time matching the colour
distribution of galaxies in the nearby Universe \citep[][]{hogg2004}. In addition, other deep spectroscopic studies in nearby galaxy clusters show that the fraction of cluster members redder than the red sequence and fainter than $M_{r} = -18.0$ is less than 10$\%$ \citep[][]{rines2006}. The spectroscopic completeness of this catalog is larger than 90$\%$ for galaxies with $m_{r} < 18.0$ and decreases to $\sim 40 \%$ at $m_{r} \sim 21.0$ \citep[][]{agulli2016b}. We have identified 460 galaxies as cluster members within the observed area. This large number of cluster members allowed us to study in detail the spectroscopic LF \citep[][]{agulli2014, agulli2016b} and the orbits of the different galaxy populations of the cluster \citep[][]{aguerri2017}.

\subsection{The large-scale structure around A\ 85}

\cite{durret1998b} showed the large-scale structure in the vicinity of A\ 85. The cluster and two galaxy groups Abell 89b (A\ 89b) and Abell 89c (A\ 89c) are located within 0.6$^{0}$ radius from the centre of A\ 85 and in the velocity range $V < 32000$ km s$^{-1}$. All these systems are observed within the field-of-view of our VIMOS and AF2  spectroscopy. The cluster and group complexes are connected by a filamentary structure visible in X-ray and optical observations \citep[][]{durret1998b}. This large scale structure suggests that the process of the mass assembly in A\ 85 is still on going. This is also suggested by the observational evidences of galaxy substructure and recent mergers in the cluster \citep[][]{ramella2007,aguerri2010, yu2016}. 

\begin{figure}
   \centering
\includegraphics[width=\hsize]{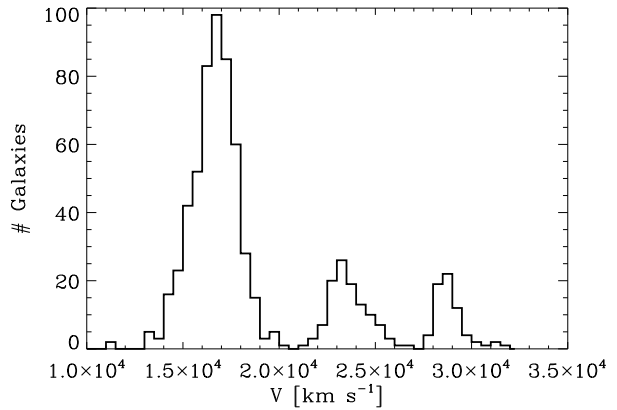}
   \caption{Velocity distribution of the galaxies in the direction of A\ 85.}
             \label{vhisto}%
    \end{figure}

Figure \ref{vhisto} shows the velocity distribution of the galaxies in the direction of A\ 85 with radial velocity smaller than 32000 km s$^{-1}$, and within 1.4 $r_{200}$. We have velocity measurements for a total of 711 galaxies in that velocity range. This sample is about two times larger than previous studies of this region of the sky \citep[][]{durret1998b}. Three peaks can be observed in the velocity histogram shown in Fig. \ref{vhisto}. The velocity peak with the largest number of galaxies corresponds to the A\ 85 cluster. This cluster is located at a mean velocity of $V_{c} = 16681$ km s$^{-1}$ and has a velocity dispersion of $\sigma_{c} = 979$ km s$^{-1}$ \citep[][]{agulli2016b}.  The other two peaks in the velocity histogram were also present in the analysis given by \cite{durret1998b}. The analysis of the dynamics of A\ 85 shows that this cluster is not relaxed. \cite{aguerri2007} showed that A\ 85 has galaxy substructure according with the Dressler-Scheckman (DS) test. Recently, \cite{agulli2016a} and \cite{yu2016} also reported the presence of substructure in this cluster. In addition, \cite{yu2016} showed that A\ 85 has a complex dynamic structure with up to four galaxy groups passing through the virial cluster region. This study shows that about $20\%$ of the galaxy cluster members of A\ 85 are located in these substructures.

The galaxies between $20000 <$ V $< 27000$ km s$^{-1}$ have a mean velocity of $<\rm{V}> = 23242$ km s$^{-1}$ and show a dispersion of $\sigma= 594$ km s$^{-1}$. As \cite{durret1998b} pointed out this velocity peak is strong in the direction of the cluster A\ 89. This cluster is located at higher redshift ($z = 0.1371$; see NED database), and following \cite{durret1998b} we will call this peak as A\ 89b. We have analyzed the gaussianity of the velocity distribution of those galaxies in the A\ 89b structure. The kurtosis and skewness of the velocity distribution are 0.57 and 0.14, respectively. These values indicate that the velocity distribution of the galaxies forming the A\ 89b structure is not Gaussian and all these galaxies do not form a single virialized structure.

The third peak shown in the velocity histogram correspond to galaxies located in the velocity range $27000 < \rm{V} < 32000$ km s$^{-1}$ \citep[][]{durret1998b}. They are located at a mean velocity of $<\rm{V}> = 28578$ km s$^{-1}$ and show a dispersion of  $\sigma = 564$ km s$^{-1}$. \cite{durret1998b} show that this peak is also in the direction of A\ 89. Following their notation, we will refer to the galaxies forming this velocity peak as A\ 89c. The analysis of the velocity distribution of the galaxies forming the A\ 89c structure shows that the skewness and kurtosis are 0.39 and -0.37, respectively. As in the case of A\ 89b, these values indicate that the velocity distribution is also not Gaussian and the galaxies of A\ 89c do not form a virialized structure. The velocity dispersions shown by A\ 89b and A\ 89c are similar and can correspond to two groups of galaxies. However, these values could be increased by the presence of substructure \citep[][]{durret1998b}.

\begin{figure*}
   \centering
\includegraphics[width=\hsize]{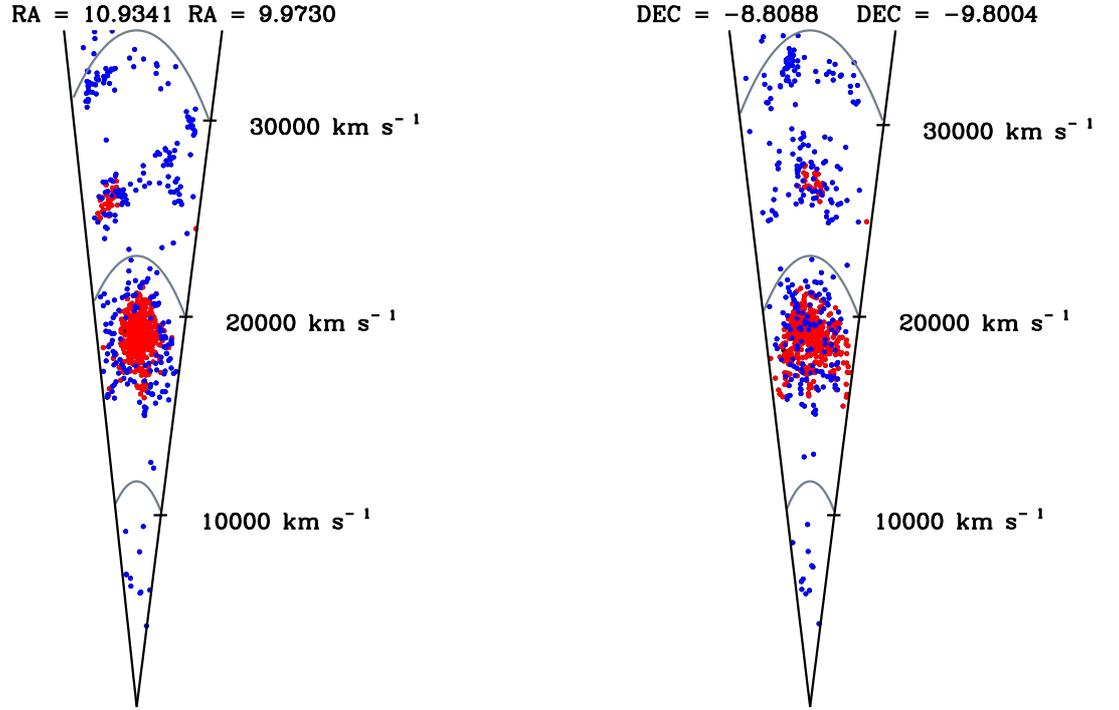}
   \caption{Spatial distribution of galaxies in the direction of Abell 85. The blue and red points represent galaxies located at local densities smaller and larger than $\Sigma_{c} \sim 30$ Gal Mpc$^{-2}$, respectively (see Sec. \ref{sec3.1} for more details). The full gray lines represent isovelocity arcs.}
             \label{lss}%
    \end{figure*}

Figure \ref{lss} shows the large scale structure around A\ 85. The highest galaxy density correspond to A\ 85. This cluster appears in Fig. \ref{lss} as a well defined and compact system. At larger velocities, galaxies are located in a complex structure formed by some groups of galaxies linked by a filamentary structure. The places of this filamentary structure with the highest densities correspond to the groups A\ 89b and A\ 89c described by \cite{durret1998b}. A\ 89b shows a more compact galaxy distribution than A\ 89c. As pointed out by \cite{durret1998b}, A\ 89b could be a galaxy group while A\ 89c could be a more complex structure formed by several groups of galaxies. 

The large-scale structure shown in Fig. \ref{lss} is similar to the observed by \cite{durret1998b}. Nevertheless, the large number of galaxies in our catalog makes more clear the different galaxy environments. Figure \ref{lss} shows that the infall region of A\ 85 has a large variety of environments. This makes it an ideal framework in order to study galaxy transformations in different environments, from  clusters  like A\ 85 to filaments connecting small galaxy groups like A\ 89b and A\ 89c.

\begin{figure}
   \centering
\includegraphics[width=\hsize,angle=-90,scale=0.7]{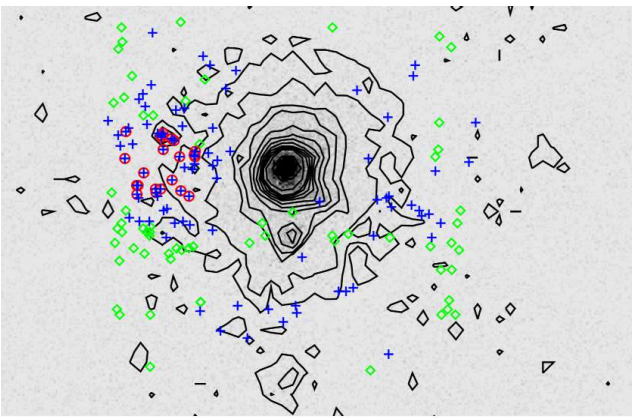}
   \caption{ROSAT image and contours overplotted from A\ 85. The blue crosses represent galaxies from the A\ 89b structure. Red circles are those galaxies from A\ 89b, but located at high galaxy density environments ($\Sigma > 40$ Gal Mpc$^{-2}$). Green diamonds are galaxies from the A\ 89c structure (see text for details).}
             \label{rosat}%
    \end{figure}

Figure \ref{rosat} shows the ROSAT X-ray image and contours of A\ 85. We have also overplotted the position of the galaxies from the A\ 89b and A\ 89c structures. Note that the galaxies from A\ 89c structure show an extended location and are not associated with any X-ray peak. Galaxies from A\ 89b structure also show an extended location. Nevertheless, those galaxies from A\ 89b at high galaxy density, $\Sigma > 40$ Gal Mpc$^{-2}$, are more concentrated. Indeed, some of them are located close to one weak X-ray peak. This indicates that some of the galaxies from A\ 89b could form a group.

    \subsection{The local galaxy density}
    
    In the present work we have analyzed several properties of the galaxies in the infall region of A\ 85 as a function of the local galaxy density and the clustercentric radius. For the later, we have adopted the centre of the cluster at the position of the brightest cluster galaxy (BCG). This corresponds to the coordinates: $\alpha_{\rm{BCG}}(\rm{J2000}) = 00^{h} 41^{m} 50^{s}.448 \  \delta_{\rm{BCG}}(\rm{J2000}) = -09^{o}18^{'} 11^{''}.45$.  This is a sensible choice because the peak of the X-ray emission of the cluster is just 7 kpc away from this centre \citep[][]{agulli2016b}. The local galaxy density was computed according to the following  procedure. 
    
We have computed the galaxy density, $\Sigma$, around each galaxy with $\rm{V} < 32000$ km s$^{-1}$ and located within 1.4$r_{200}$. This density was obtained assuming a fix aperture of 500 kpc and a velocity difference of $\pm 1000$ km s$^{-1}$ around each galaxy.  The spectroscopic completeness of our catalog was taken into account. Thus, the galaxy density around the $i-th$ galaxy is given by:

\begin{equation}
 \Sigma_{i} = N_{i,gal}/A,
 \end{equation}
 
where $N_{i,gal}$ was computed by:

\begin{equation}
N_{i,gal} = N_{i,gal,vel} + N_{i,phot} \times \frac{N_{i,gal,vel}}{N_{i,vel}}
\end{equation}

$N_{i,gal,vel}$ is the number of galaxies within 500 kpc and $\pm 1000$ km s$^{-1}$ around the $i-th$ galaxy. $N_{i,phot}$ is the number of photometric targets of the spectroscopic survey with no velocity information and within 500 kpc around the $i-th$ galaxy. $N_{i,vel}$ is the number of galaxies with measured velocity within 500 kpc around the $i-th$ galaxy. The second term of Eq. (2) takes into account the spectroscopic completeness of the survey \citep[][]{agulli2014,agulli2016b}. In Eq. (1), A corresponds to the area in Mpc$^{2}$ considered around each galaxy. This area corresponds to $\pi \times 0.5^{2}$ Mpc$^{2}$. Edge effects were taken into account for those galaxies located close to the limit of the survey. We have also explored aperture effects on the galaxy density. No significant variations in the relations related with the local galaxy density  were obtained by using a fix aperture of 250 kpc.

\subsection{Spectral classification of the galaxies}
\label {sec2.4}

The galaxies with $\rm{V}<32000$ km s$^{-1}$ and with data from SDSS-DR13 (212 spectra) and VIMOS@VLT (432 spectra) were classified according to their spectral features. In particular, we have used for this classification the equivalent width (EW) of the O[II]\ $\lambda 3727$ \AA\ and H$\delta\  \lambda 4102$ \AA\ spectral lines. The data from AF2@WHT were not used for the spectral classification because the [OII]$\lambda 3727 \AA$ line is out of the wavelength range of these observations. This AF2 data represent 9$\%$ of the total spectroscopic sample.

The EW of O[II] and H$\delta$ lines were measured following the definitions given by \cite{balogh1999}.  We have adopted the convention of identifying emission and absorption lines with negative and positive EWs, respectively.

The galaxies were classified in three groups according to their [OII] and H$\delta$ EWs. We have called emission line (EL) galaxies those with EW([OII])$ < -5$ \AA. PSB galaxies are those showing  EW(H$\delta) > 5$ \AA\ and EW([OII])$ > -5$ \AA. The third group is formed by passive (PAS) galaxies which have EW(H$\delta) < 5$ \AA\ and EW([OII])$ > -5$ \AA, similar to the adopted by \cite{balogh1999}. We have visually checked the individual spectra of the galaxies in order to see that those galaxies with EW([OII])$ > -5$  \AA\ do not have emission lines. The limit of EW(H$\delta) = 5$ \AA\ is the same as the adopted in other studies of PSB galaxies \citep[see e.g.,][]{balogh1999, goto2007}. Other surveys use values of EW(H$\delta$) = 3\ \AA \ to select PSB galaxies \citep[e.g.,][]{paccagnella2017}. The main difference with our selection would be that we are selecting PSB galaxies that have stopped their star formation more recently \citep[see e.g.,][]{goto2004}.

\begin{figure}
   \centering
\includegraphics[width=\hsize]{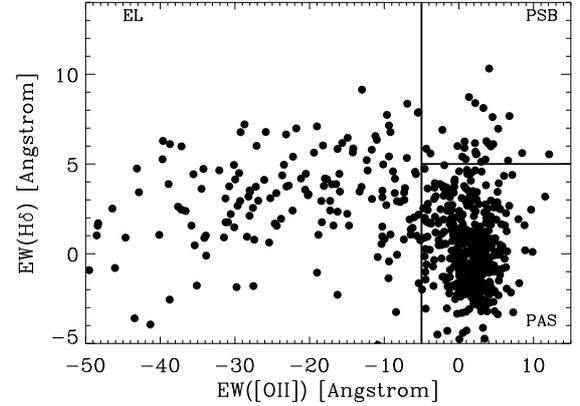}
   \caption{H$\delta$ and [OII] EWs of the galaxies with V < 32000 km s$^{-1}$. The vertical and horizontal lines show the limits of EL, PSB, and PAS galaxies (see Sec. \ref{sec2.4} for more details).}
             \label{specclas}%
    \end{figure}

Figure \ref{specclas} shows the location of each galaxy in the plane EW(H$\delta$) - EW([OII]). According to this, we have 68.5$\%$, 27.0$\%$, and 4.5$\%$  of PAS, EL, and PSB galaxies, respectively. For the cluster members, we have 78.8$\%$, 16.0$\%$ and 5.2$\%$  of PAS, EL, and PSB galaxies, respectively. The percentages for those galaxies non cluster members are: 41.0$\%$, 56.0$\%$, and 3.0$\%$ for PAS, EL, and PSB galaxies, respectively. Notice that the percentage of PAS galaxies is larger within the cluster population. In contrast, the fraction of EL is larger for the non cluster members. This result reflects the well known morphology density relation \citep[e.g.,][]{dressler1980, dressler1997} or the star formation - density relation \citep[e.g.,][]{haines2015}. The percentage of PSB galaxies changes $\sim 2\%$ from cluster to non-cluster members, being larger in the cluster population. Due to our small number of PSB galaxies, we can not consider this variation in the fraction of PSB significant.

\begin{figure}
   \centering
\includegraphics[width=\hsize]{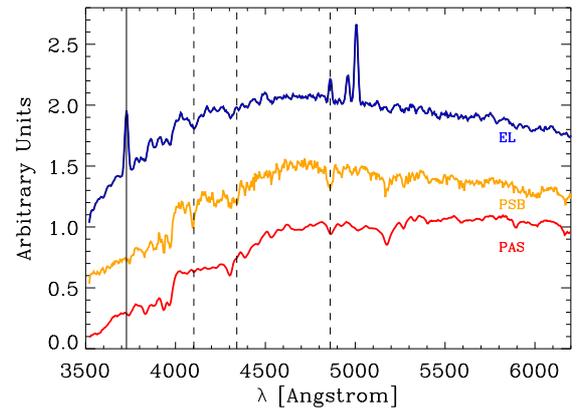}
   \caption{Rest-frame stacked spectra of PAS (red), PSB (orange) and EL (blue) galaxies. The vertical dashed lines show the wavelengths (from larger to shorter wavelengths) of the H$\beta$, H$\delta$ and H$\gamma$ Balmer lines. The vertical full line represents the [OII]$\lambda 3727\  \AA$ emission line.}
             \label{specstack}%
    \end{figure}

Figure \ref{specstack} shows the stacked rest-frame spectra of PAS, EL, and PSB galaxies. The spectra were stacked by normalizing their flux at 5100\ \AA. The PSB spectrum shows strong absorption at the Balmer (H$\beta$, H$\delta$, and H$\gamma$) lines and no emission at the [OII] line. The spectrum of the PAS galaxies has no emission at the [OII] line but the absorption at the Balmer lines is less strong. The spectrum of the EL galaxies indicates active star formation as it is evident from the emissions lines at [OII], H$\beta\ \lambda$4861\ \AA\ and the [OIII] $\lambda 4996, 5007$ \AA\ doublet.

   \section{Results}
    
        \subsection{First hint of quenching: Galaxy density and stellar colour relation}
    	\label{sec3.1}

\begin{figure}
   \centering
\includegraphics[width=\hsize]{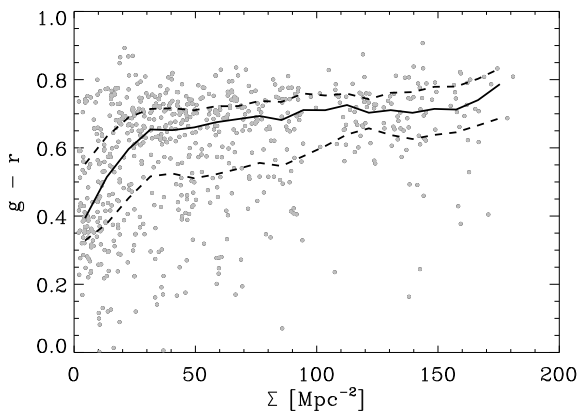}
   \caption{Colour - density diagram of the galaxies with $V < 32000$ km s$^{-1}$ in the direction of A\ 85. The full black line represents the quartile of the 50$\%$ of the distribution. The dashed lines shows the 25$\%$ and 75$\%$ quartiles.}
             \label{fcolor}%
    \end{figure}

Figure \ref{fcolor} shows the $g - r$ stellar colour of the galaxies with $\rm{V} < 32000$ km s$^{-1}$ as a function of the local galaxy density. Galaxies are bluer in less dense environments and redder in high density ones. This relation have been shown in other galaxy samples before \citep[e.g.,][]{kodama2001,sanchezjanssen2008}. Our relation is based on spectroscopic velocities down to the dwarf galaxies ($\sim M_{r}^{*} +6$) not always observed in other surveys. The dependence of the galaxy stellar colour with the local density suggests a quenching of the galaxies from low to high galaxy densities. 

There is a critical density ($\Sigma_{c} \sim 30$ Gal Mpc$^{-2}$) above which there is almost no change in the mean stellar colour of the galaxies. In contrast, galaxies at $\Sigma < \Sigma_{c}$ show a strong correlation between their $g -r$ colour and $\Sigma$.  \cite{kodama2001} also reported this colour-density relation for their galaxies in the medium redshift cluster Abell 851. They also found a critical density where the transition from blue to red colour occurs. The change in colour from low to high density was explained as a decrease of the star formation rate (SFR) of the galaxies by a factor 6.

We have colour coded the galaxies in Fig. \ref{lss} according to this critical galaxy density. Thus, galaxies in environments with $\Sigma < \Sigma_{c}$ are shown as blue points in Fig. \ref{lss}. They are located in the external regions of A\ 85 and in the filamentary structure located outside the cluster. In addition, these galaxies are also located in the A\ 89c group (see Fig. \ref{lss}). This indicates that the A\ 89c group  is probably not as massive as its velocity dispersion indicates. As suggested by \cite{durret1998b}, A\ 89c group could be a complex structure formed by several small groups and substructures given the large value of its velocity dispersion.   Galaxies in environments with $\Sigma > \Sigma_{c}$ are shown with red symbols in Fig. \ref{lss}. These galaxies are located in the A\ 85 and A\ 89b environments. The fact that  A\ 89b has galaxies located at densities higher than $\Sigma_{c}$ indicates that this is a rich group in agreement with its velocity dispersion.

\begin{figure}
   \centering
\includegraphics[width=\hsize]{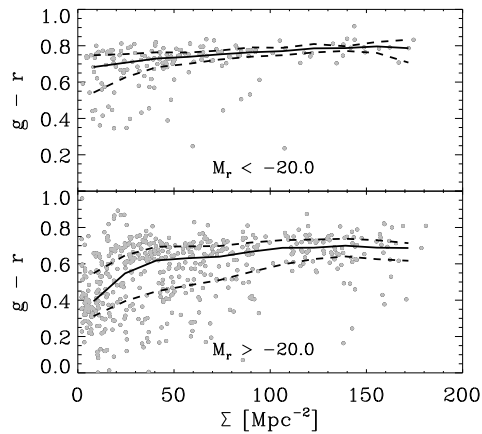}
   \caption{Colour - density diagram of the galaxies in the direction of A\ 85 with  $M_{r} < -20.0$ (top panel) and $M_{r} > -20.0$ (bottom panel).  In both panels, the full black line represents the quartile of the 50$\%$ of the distribution. The dashed lines shows the 25$\%$ and 75$\%$ quartiles.}
             \label{fcolormag}%
    \end{figure}

Figure \ref{fcolormag} shows the colour-density relation for those galaxies with $M_{r} < -20.0$ and $M_{r} > -20.0$. There is no relation between the $g - r$ stellar colour and the local galaxy density for those galaxies with $M_{r} < -20.0$. These galaxies show a colour $g - r \sim 0.7$ independently of the local density. Similar stellar colour in the bright galaxy population have been observed in other nearby galaxy clusters \citep[e.g.,][]{agulli2016a}. In contrast, faint galaxies ($M_{r} > -20.0$) show a clear colour segregation. Those faint objects located at $\Sigma < \Sigma_{c}$ are bluer than those at larger galaxy densities. This means that the abrupt change in colour observed in Fig. \ref{fcolor} is mainly due to galaxies with $M_{r} > -20.0$.  

The mass-to-light (M/L) ratio in the $r$-band for a galaxy with $g - r =0.7$ is 2.9 \citep[see][]{bell2003}. This means that a galaxy with $M_{r} = -20.0$ has a mass of $\sim 2  \times 10^{10}$ M$_{\odot}$. Thus, the non-dependence of the stellar colour and the galaxy density for galaxies brighter than $M_{r} = -20.0$ is in agreement with the results obtained by \cite{peng2010}. They suggested that 
the quench of galaxies more massive than $10^{10}$ M$_{\odot}$ is dominated by internal galaxy processes (self quench). In contrast, less massive objects show a quenching dominated by environmental processes (environmental quenching).  Another explanation for this colour-luminosity dependence could be that more luminous (massive) galaxies in clusters and groups, as those observed here, are already evolved objects. This means that the effect of the environment, if any, would have happened long time ago in their formation history

The colour - density relation shows a first hint about the quenching of the galaxies around A\ 85. We will focus now on the properties of the galaxies that have been recently quenched (PSB galaxies) in order to understand the mechanisms driven this transformation.

    \subsection{Properties of PSB galaxies.}

We present in this section the main physical properties (absolute $r$-band magnitude, $g - r$ stellar colour, and local galaxy density) of the PAS, EL, and PSB galaxies with $V < 32000$ km s$^{-1}$. We have divided the galaxies in three different samples: i) the full sample contains all galaxies with spectral classification within $V<32000$ km s $^{-1}$ and within 1.4 $r_{200}$; ii) the cluster member sample is formed by those galaxies selected as cluster members of A\ 85 \citep[][]{agulli2016b}; and iii) the non-cluster member population is formed by those galaxies not selected as cluster members of A\ 85. This last galaxy population contains the galaxies located in the A\ 89b and A\ 89c groups and in the filamentary structure connecting these groups.

\begin{figure*}
   \centering
\includegraphics[width=\hsize,height=10cm]{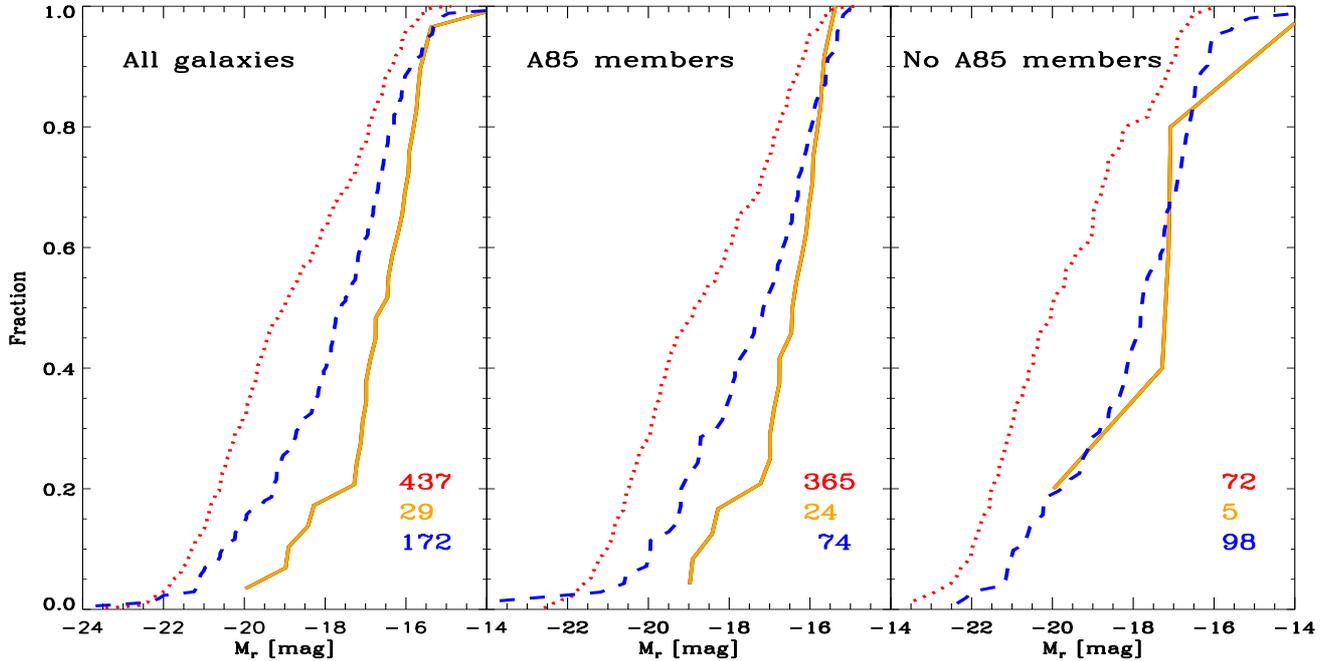}
   \caption{CDFs of $M_{r}$ for PAS (red dotted line), EL (blue dashed line) and PSB (orange solid line) galaxies for all galaxies (left panel), members of A85 (middle panel), and non A85 members (right panel). The red, orange and blue figures show the number of PAS, PSB and EL galaxies, respectively.}
             \label{magpsb}%
    \end{figure*}

Figure \ref{magpsb} shows the cumulative distribution function (CDF) of the absolute $r$-band magnitudes ($M_{r}$) for  EL, PAS, and PSB galaxies corresponding to the three previously defined populations of galaxies. For all galaxy samples, PAS galaxies are, on average, brighter than EL and PSB galaxies. The PSB galaxies are the faintest ones for the full and cluster members samples. Nevertheless, they show similar $M_{r}$ than EL galaxies for the non-cluster members sample. 

All galaxy samples show that most of the PSB galaxies ($\sim 70 - 80\%$) are dwarfs (M$_{r} > -18.0$).  This indicates that, independently of the environment, the galaxies that  have recently stopped their star formation in A\ 85 and its infall region are mainly dwarf galaxies.  This result was also reported in other samples of PSB galaxies \citep[e.g.,][]{dressler2013, cybulski2014, paccagnella2017}. We did not observe PSB galaxies brighter than $M_{r} =-20.0$. 

We have computed the stellar mass of the galaxies by using their $g - r$ colour and following the prescription given by \cite{zibetti2009}. We have also observed a clear mass segregation. Thus, no PSB galaxies were observed with masses larger than 6.05$ \times 10^{9} M_{\odot}$. In addition, 75$\%$ of the PSB galaxies show stellar masses smaller than $10^{9} M_{\odot}$. These findings confirms that the today PSB phase in A\ 85 is related with dwarf galaxies.

\begin{figure*}
  \centering
\includegraphics[width=\hsize,height=10cm]{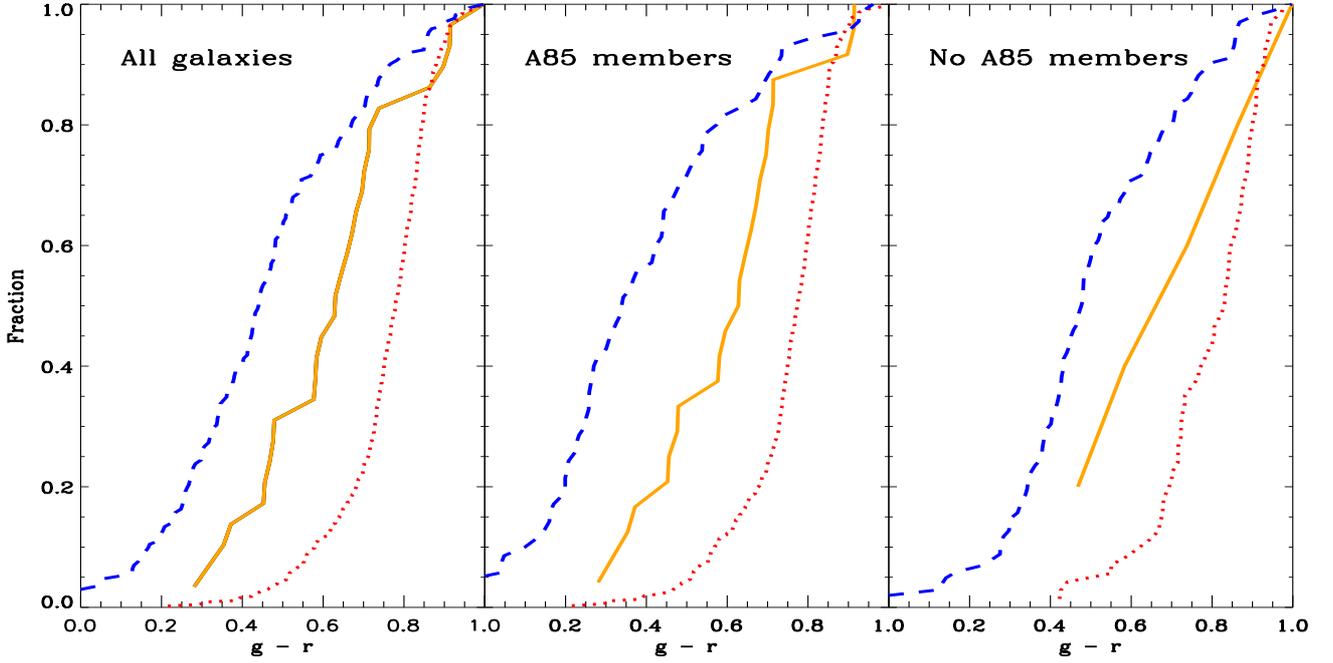}
   \caption{CDFs of $g -r$ stellar colour for PAS (red dotted line), EL (blue dashed lin) and PSB (orange solid line) galaxies for all galaxies (left panel), members of A85 (middle panel), and non A85 members (right panel). The number of galaxies in each panel as in Fig. \ref{magpsb}}
             \label{colorpsb}%
    \end{figure*}

Figure \ref{colorpsb} shows the CDFs of the $g - r$ stellar colour for EL, PAS and PSB galaxies. For all galaxy samples, PAS galaxies show the reddest $g-r$ colour while EL have the bluest ones. PSB galaxies show $g-r$ stellar colour between PAS and EL galaxies. The mean $g - r$ color of PSB, PAS and EL galaxies is 0.6, 0.8 and 0.4, respectively. These mean colors shows that PSB are mainly located in the so-called green valley of the CMD \citep[e.g., ][]{strateva2001,baldry2004,lackner2012,mendez2011}. They would be objects in transition between  the blue cloud and the red sequence.
    
\begin{figure*}
   \centering
\includegraphics[width=\hsize,height=10cm]{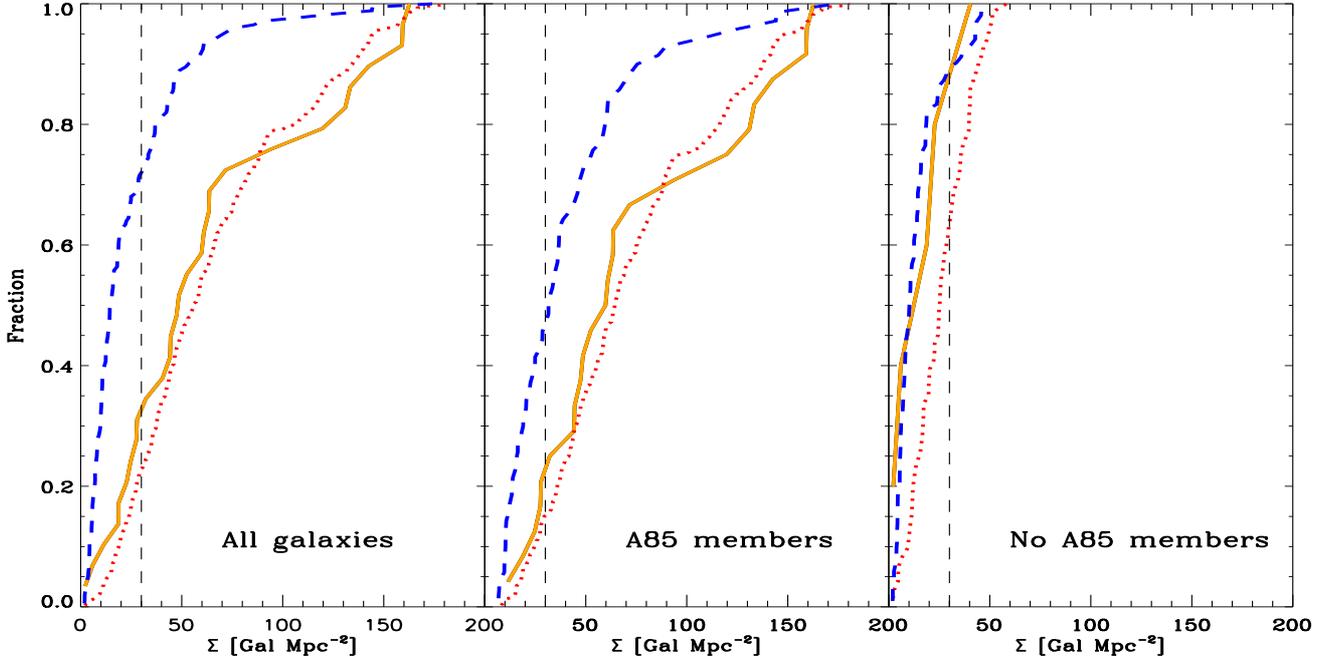}
   \caption{CDFs of local galaxy density for PAS (red dotted line), EL (blue dashec line), and PSB (orange solid line) objects for all galaxies (left panel), A\ 85 members (central panel), and non-cluster members (right panel). The vertical dashed lines represent $\Sigma_{c}$ (see text for more details). The number of galaxies in each panel as in Fig. \ref{magpsb}}
             \label{denpsb}%
    \end{figure*}

Figure \ref{denpsb} shows the CDFs of the local galaxy density for EL, PAS, and PSB galaxies. Regardless of the environment, EL galaxies are located at lower local densities than PAS ones. PSB galaxies share similar local densities as PAS ones for the cluster member sample. In contrast, PSB galaxies show similar galaxy densities as EL ones in the non-cluster member population. This was confirmed by the results of the Kolmogorov-Smirnoff (KS) test. 

The results obtained from the luminosity and local density distributions for PSB galaxies suggest that there is a link between EL and PSB galaxies for the non-cluster members.

\begin{figure*}
   \centering
\includegraphics[width=\hsize]{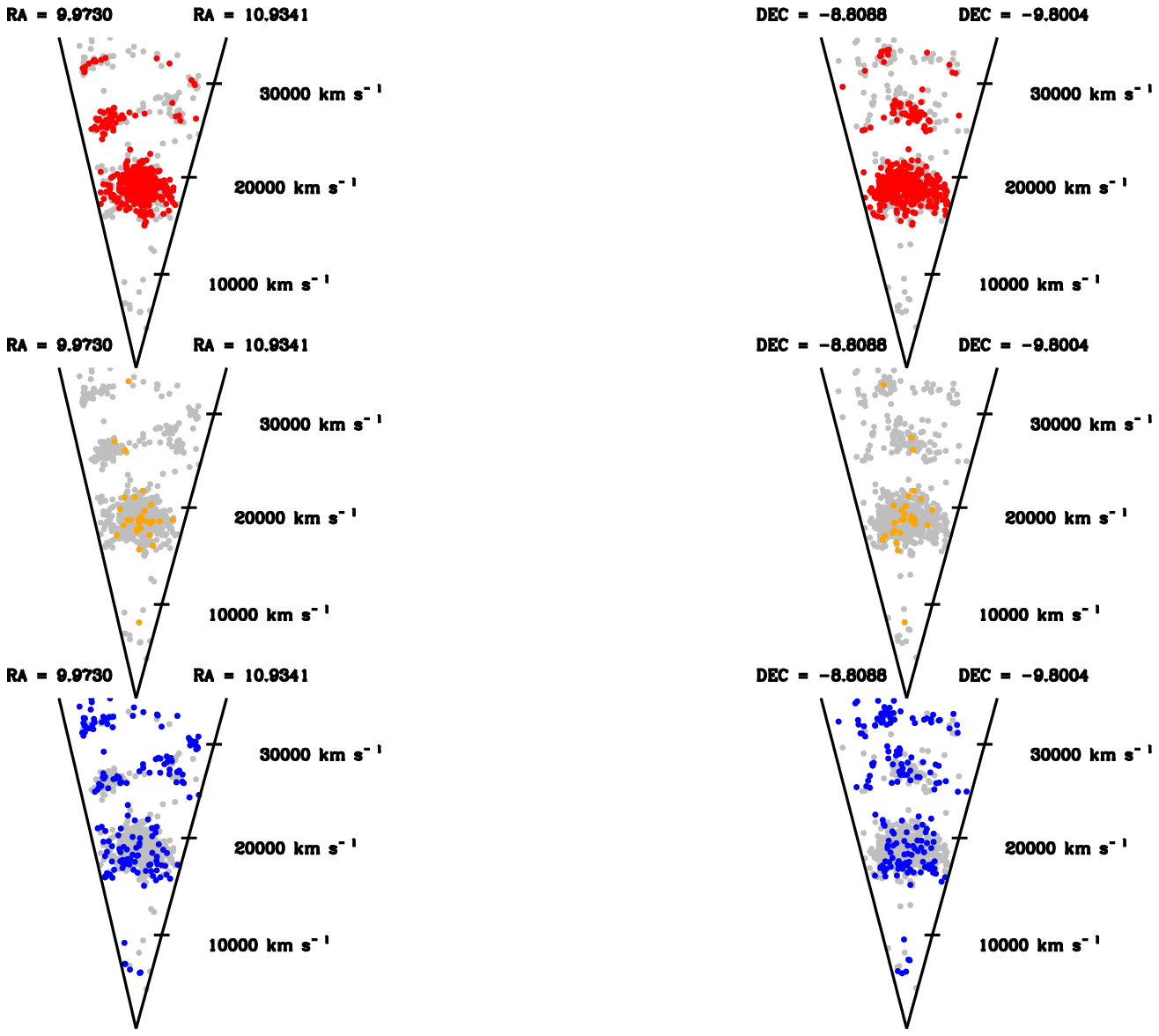}
   \caption{Spatial distribution of galaxies in the direction of Abell 85.  They grey dots indicates those galaxies with redshift. The red, orange and blue dots correspond to PAS, PSB, and EL galaxies, respectively.}
             \label{lss_types}%
    \end{figure*}

Figure \ref{lss_types} shows the location of the different spectral types of galaxies in the different structures. PAS galaxies are located in places of large local galaxy density (see also Fig. \ref{lss}). In contrast, EL galaxies are located in less dense environments, i.e., mainly in the most external regions of A\ 85 and in the A\ 89b and A\ 89c structures. Indeed, the A\ 89c structure is dominated by EL galaxies.  and only 23$\%$ of the galaxies are classified as PAS ones. This is another indication that this structure is formed by a complex of galaxies located in small groups and filaments. The PSB population is mainly located in A\ 85. Only 4 PSBs are located outside of the A\ 85 environment. Two of them are located close to the highest density region of A\ 89b. The remaining two are located in lower galaxy density environments. This indicates that PSB galaxies are located in all different environments (cluster, groups and filaments) that we observed around A\ 85.

 \subsection{Dynamics and location of the different spectral types inside A\ 85}
 
\begin{figure*}
   \centering
\includegraphics[width=\hsize]{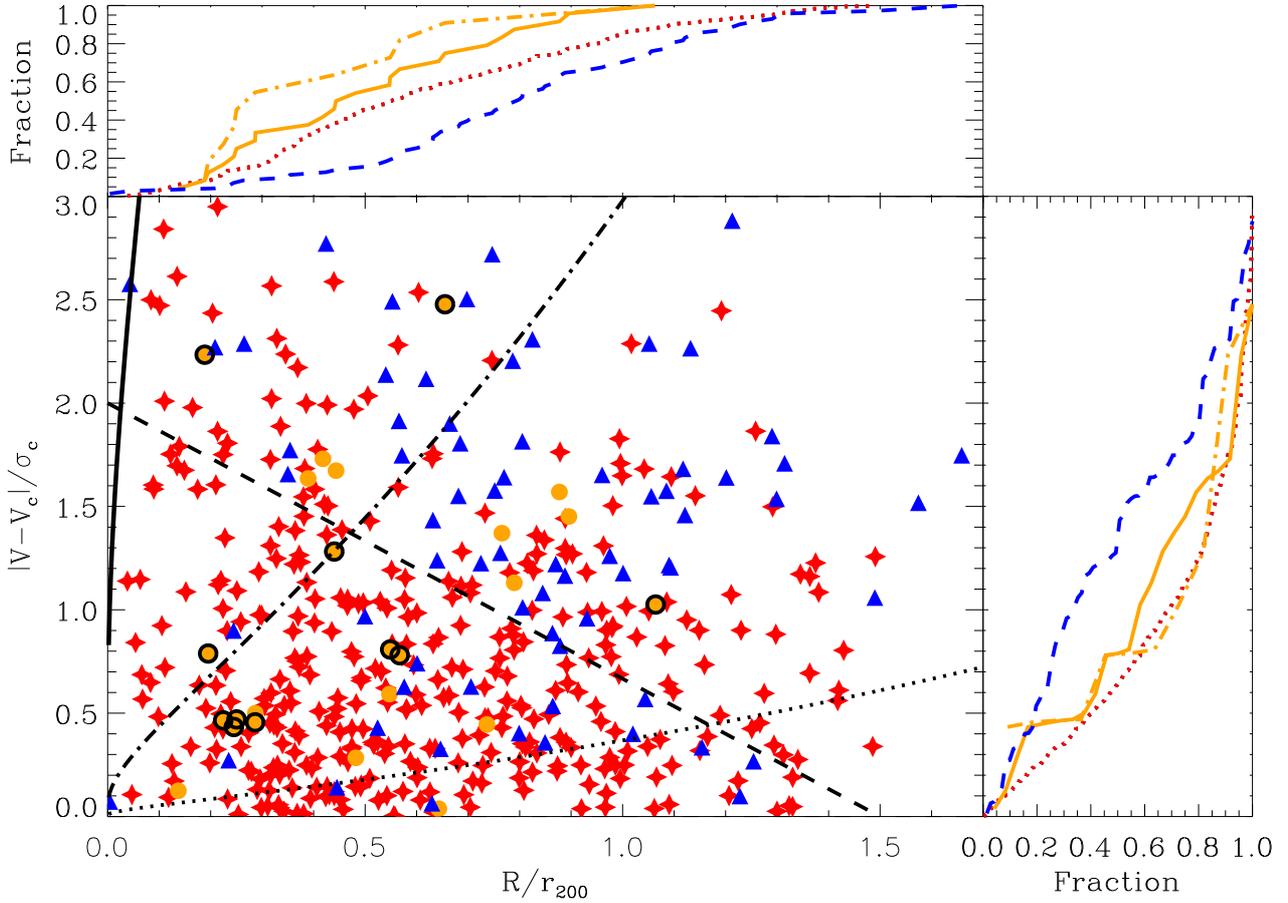}
   \caption{(Central panel) Phase-Space diagram of the galaxies in A\ 85. The red stars, blue triangles and orange filled points represent  PAS, EL and PSB galaxies, respectively. The black empty circles represent the sPSB galaxies. The dashed line separates the infall and virialized regions of the cluster according to \citet{oman2013}. The full, dot-dashed and dotted lines delimites the regions (to the left of the lines) were galaxies of $10^{10}, 10^{9}$ and $10^{8}$ M$_{\odot}$ would be stripped by ram-pressure, respectively. See more details about this lines in the discussion section. (Upper panel) CDFs of $R/r_{200}$ for the PAS (red dotted line), PSB (orange solid line), sPSB (orange dot-dashed line), and EL (blue dashed line) cluster members. (Right panel) CDF of $|V - V_{c}|/ \sigma_{c}$ for the PAS (red dotted line), PSB (orange  solid line), sPSB (orange dot-dashed line), and EL (blue dashed line) cluster members. }
             \label{velpsb}%
    \end{figure*}

Figure \ref{velpsb} shows the the phase-space diagram of the PAS, PSB, and EL cluster member galaxies. In addition, this figure also shows the  CDFs of $(|V-V_{c}|)/\sigma_{c}$ and $R/r_{200}$ for those galaxies. The KS test reports that the CDFs of EL and PAS galaxies are statistically different. In particular, PAS galaxies show smaller values of $(|V-V_{c}|)/\sigma_{c}$ and $R/r_{200}$ than EL ones. This velocity and location segregation between PAS (red) and EL (blue) galaxies have been found in other studies about galaxy clusters in the literature \citep[e.g.,][]{goto2005, sanchezjanssen2008}. 
 
The PSB  and PAS populations of A\ 85 have similar dynamics. The KS test gives that the CDFs of $(|V-V_{c}|)/\sigma_{c}$ of PSB and PAS galaxies are statistically similar. Nevertheless,  the CDFs of $R/r_{200}$ for PSB and PAS galaxies are statistically different.  Thus, the PSB population is located at smaller distances from the cluster centre than the PAS population. 

\cite{paccagnella2017} show that the velocity dispersion profile of the PSB population with EW(H$\delta$) < 6 $\AA$ is similar to the PAS galaxies in a sample of nearby galaxy clusters. In contrast, the velocity dispersion profile of the strongest PSB (sPSB) galaxies (EW(H$\delta > 6 \AA$) is similar to the EL ones. The strong absorption H$\delta$ line shown by the sPSB population can be interpreted as a high mass burst of star formation happening prior to the sudden quench \citep[][]{paccagnella2017}. These objects are PSB galaxies that have been caught in a very early phase of the transition \citep[][]{goto2004}. \cite{paccagnella2017} interpreted the different dynamics of PSB and sPSB galaxies as if they are in different accretion states.

 We find 11 sPSB galaxies between the cluster member population. The CDFs of $(|V-V_{c}|)/\sigma_{c}$ for these galaxies in A\ 85 is also shown in Fig. \ref{velpsb}. In this case, there is no difference between the CDFs of PAS and sPSB samples. In the case of A\ 85 the populations of PSB and sPSB galaxies show similar dynamics. However, the CDF of $R/r_{200}$ shows that the sPSB galaxies are the population located closest to the cluster centre (see Fig. \ref{velpsb}). This indicates that there is a correlation between the strength of the EW(H$\delta$) and the position in the cluster. This point towards a link between the quench of the star formation of the galaxies and their position within the cluster.
 
\cite{oman2013} showed that the position of the cluster galaxies in the phase-space diagram depends on their infall time into the cluster. They proposed that the phase-space can be divided using the line: $|V-V_{c}|/\sigma_{c} = -\frac{4}{3} \frac{R}{r_{200}} +2.0$. Then, for a fixed clustercentric radius, galaxies in the phase space above this line fell into the cluster during the last 1 Gyr. In contrast, those galaxies below the line fell into the cluster several Gyr ago and they are located in the virialized cluster region. We have plotted this line in the phase-space diagram of A\ 85 (see Fig. \ref{velpsb}). Galaxies in the virialized cluster region are mostly PAS ones. In contrast, the infall region is dominated by the EL galaxies.  PSB galaxies are located both in the virialized (60$\%$) and the infall (40$\%$) regions. 

\begin{figure}
   \centering
\includegraphics[width=\hsize]{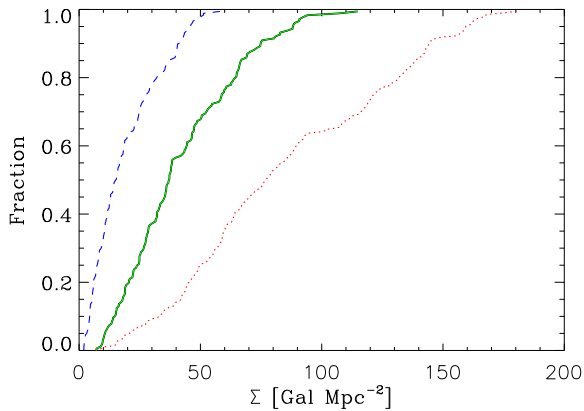}
   \caption{CDFs of $\Sigma$ for galaxies in the virialized (red dotted line) and infall (green solid line) cluster regions. The blue dashed line represents the CDF of $\Sigma$ for those galaxies non cluster members. }
             \label{phaceden}%
    \end{figure}

We have analyzed the local density of those galaxies located in the virialized and the infall cluster regions. This galaxy density has been compared with that of galaxies classified as non-cluster members, i. e., galaxies with $V > 20000$ km s$^{-1}$ (see Fig. \ref{phaceden}). Notice that galaxies in the infall cluster region are located at larger galaxy densities than non-cluster members. This indicates that galaxies in the infall cluster region of A\ 85 suffer stronger environments than the non-cluster member galaxies (i.e. those galaxies located in the complexes A\ 89b and A\ 89c).

\subsection{Morphology of the galaxies for the different spectral types}

The plane defined by the galaxy effective radius ($r_{e}$) and its absolute $r$-band magnitude ($M_{r}$) could be used in order to have a qualitative classification of the galaxies in two broad morphological groups: early- and disc-type galaxies. This is due to disc-type and early-type galaxies are located at different regions of this plane \citep[e.g.,][]{shen2003}. Thus, for a given $M_{r}$ early-type galaxies are more compact (smaller $r_{e}$) than disc-type objects. This is a broad galaxy classification. One strength of this classification is that  was done until the same limiting magnitude that our galaxy sample \citep[][]{shen2003}. We have used this relation to classify the galaxies of our database in these two broad morphological groups. Both $r_{e}$ and $m_{r}$ of the galaxies were obtained from the SDSS database.

\begin{figure}
\centering
\includegraphics[width=10cm]{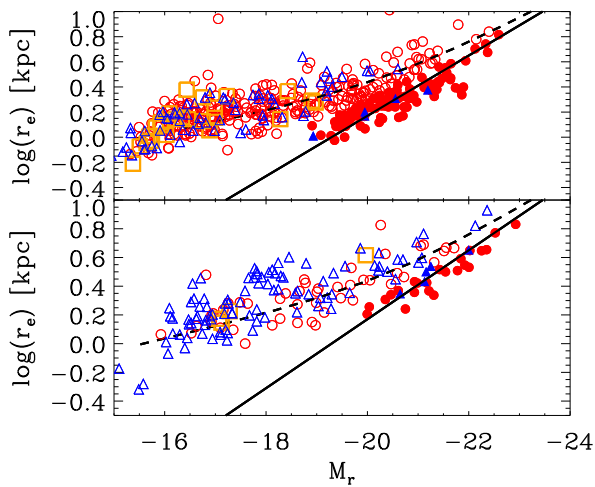}
   \caption{Magnitude-size relation for the cluster (top panel) and non-cluster  member galaxies (bottom panel) for  EL (blue triangles), PSB (orange squares), and PAS (red circles) objects. The full and dashed lines \citep[][]{shen2003} show the location in the $M_{r} - r_{e}$ plane of the early- and disc-types galaxies, respectively. Filled and empty symbols represent whether the galaxy is closer to the relations of the early-type of disc-type, respectively.}
 \label{magre}%
 \end{figure}

Figure \ref{magre} shows the location of the cluster and non-cluster members in the $M_{r} - r_{e}$ plane. The separation between the two morphological groups was done by using the distance of each galaxy to the relations  given by \cite{shen2003} for the disc-type and early-type galaxies. According to this classification, we have obtained that 80$\%$ and 20$\%$ of the full galaxy population are disc-type and early-type objects, respectively. These percentages  change when the cluster and non-cluster members samples are considered. Thus, 22$\%$ and 78$\%$ are classified as early-type and disc-type galaxies among the cluster members. For the non-cluster galaxies, 16$\%$ and 84$\%$  are classified as early-type and disc-type objects, respectively. The large fraction of disc-type galaxies independently of the environment is due to all dwarf galaxies are classified as disc-like objects. There are not compact dwarf ellipticals (cE) in the A\ 85 environment as observed in other nearby clusters environments \citep[e.g.,][]{kormendy2009}.

The early-type galaxies are dominated by PAS ones (92$\%$). Only 8$\%$ of the early-type galaxies are classified as EL.  The early-type population are mainly cluster members (81$\%$). The disc-type population is formed by PAS (62$\%$), PSB (6$\%$) and EL (32$\%$) galaxies. In this case, the non-cluster disc-type population is dominated by ELs (63$\%$). In contrast, the largest fraction of disc-type cluster member galaxies are those with PAS (74$\%$) spectral classification. The PSB galaxies are always classified as disc-type for those cluster and non-cluster members.

\section{Discussion}

\subsection{The mechanisms driving the quenching of galaxies in the infall region of A\ 85}
The properties of the PSB galaxies presented in this study vary with the environment. For the cluster member sample, PAS and PSB galaxies are located in similar local galaxy densities and have statistically similar CDF of $|V-V_{c}|/\sigma_{c}$. In contrast, PSB galaxies outside A\ 85 shows similar magnitudes and local galaxy densities than EL ones. These differences in the properties of the PSB galaxies with environment might indicate that the recent quench of the their star formation has been produced by different mechanisms. 

\subsubsection{Formation of PSB in the non-cluster member sample}

The fact that PSB and EL galaxies in the non-cluster member sample are located in similar galaxy environments suggests that they are linked. In addition, PSB and EL galaxies of the non-cluster member sample have similar $r$-band absolute magnitudes. These findings indicate that the progenitors of the PSB galaxies in the non-cluster member sample could be EL galaxies that have recently stopped their star formation. In addition, non-cluster member galaxies are located in lower density environments than galaxies located in the virialized and infall regions of A\ 85. Numerical simulations show that hydrodynamical processes are efficient in the innermost regions of the massive clusters \citep[see][]{jaffe2015}. Thus, different mechanisms should be considered in this case.

Numerical simulations suggest that major gas-rich mergers can induce starburst strong enough to rapidly consume the gas of the galaxies and drive them into a PSB phase \citep[e.g.,][]{barnes1991, bekki2005, hopkins2006, hopkins2008, snyder2011}. These interactions can be produced in similar environments as those ocupated by the PSB non-cluster member galaxies. The link between mergers and PSB galaxies have been investigated in the past \citep[][]{wild2009}. Thus, the fact that PSB have been discovered in low density environments in several surveys have pointed out to this origin \citep[e.g.,][]{zabludoff1996, quintero2004, balogh2005}. Major merger events could strongly affect the morphology of the galaxies. Thus, they can produce disturbances or asymetries in the external parts of the galaxies. This kind of morphological disturbances have been discovered in some PSB galaxies  \citep[see][]{blake2004, goto2005, yang2008, pawlik2016}. In addition, the final products of these major merger events would be galaxies with prominent spheroidal components \citep[e.g.,][]{toomre1972, barnes1988, naab2003}. Nevertheless, not all PSB galaxies show these disturbances and the number of PSBs is larger than the number of major merger rate \citep[][]{wild2009}.  Moreover, our morphological classification shows that our PSB galaxies are similar to disc-like systems. This could indicate that other mechanisms rather than major mergers could be responsible for the observed PSB galaxies.

Other possible scenarios about the formation of PSB galaxies in low density environments have been proposed by \cite{dressler2013}. They  suggested that minor mergers or accretions might be the responsible for the production of PSB galaxies. In this framework, the morphology of the main galaxy does not change significantly. In particular, minor mergers or small accretions onto disc galaxies would preserve the disc-like structure of the main galaxy \citep[e.g.,][]{aguerri2001, elichemoral2006}. This gas-rich minor merger events could activate massive starburst and produce a PSB galaxy once they are quenched. These minor events could occur more than once in the lifetime of the galaxies, and therefore they would undergo several PSB phases as well \citep[][]{dressler2013}

The stop of the star formation in low density environments could be produced by mechanisms like AGN activity or SNe feedback. Some enhance in the fraction of AGN in galaxy groups have been observed in the past \citep[][]{coziol2004, popesso2006b, arnold2009}. The type of interactions taking place in low density environments  could produce the tunneling of gas towards the central region of the galaxies, fueling the AGN activity \citep[e.g.,][]{hwang2012}. In addition, SNe feedback could also spell out the gas from galaxies and stop their star formation. This mechanism is efficient in low-mass galaxies like the PSB galaxies observed in the low density regions around A\ 85 \citep[][]{dekel1986}.

\subsubsection{Formation of PSB galaxies inside A\ 85}

The high relative velocity between galaxies in massive galaxy clusters like A\ 85 produce that galaxy mergers are less frequent than in the field or in group environments. In this case, the quench of the star formation in the galaxies inside A\ 85 is unlikely to be produced by mergers or accretions. Nevertheless, fast galaxy-galaxy interactions or interactions with the cluster potential are frequent in massive clusters and produce the loss of gas and stars stopping the star formation. In addition, the presence of a hot intracluster medium in massive clusters produces the interaction of the galaxies with this hot gas. This interaction also produces the loss of gas in the galaxies and a subsequent quench of their star formation. 

The PSB population inside A\ 85 are located at similar local galaxy densities and show similar dynamics than PAS galaxies. Nevertheless, PAS are unlikely to be the progenitors of PSB galaxies, since in general,  gas accretion or a merger with gas should be invoqued in order to activate the PAS galaxy, produce an starburst, and then pass through the PSB phase once the gas is consumed. As mentioned before mergers or accretions inside massive clusters are not frequent. Therefore, the progenitor of the PSB galaxies inside A\ 85 should also be EL galaxies. These EL galaxies could be objects recently accreted to the cluster. The typical star formation history of dwarf galaxies is extended during several Gyrs, having enough time to arrive as EL galaxies to the vicinity of the center of the cluster. They can also be galaxies with gas that activate their star formation when they are falling into the cluster potential well \citep[][]{dressler2013}. But the quench of their star formation must be produced by different physical mechanisms than in the field environment.

The mechanisms driving the quenching of the star formation in galaxies inside A\ 85 should preserve their morphology. This is because all PSB galaxies observed inside A\ 85 are compatible with being disc-like objects. Both, galaxy harassment and interactions with the cluster potential can produce strong morphological transformations in galaxies \citep[][]{mastropietro2005, aguerri2009}. This indicates that these strong gravitational interactions of the galaxies are not likely the responsible of the production of PSB galaxies inside A\ 85. Hydrodynamical mechanisms like ram-pressure stripping or starvation can produce the quench of the star formation in galaxy clusters and do not change the morphology of the objects.

Both ram-pressure stripping and starvation produce the lost of gas in galaxies and the subsequent quench of the star formation. Nevertheless, the quenching time-scale is very different for these two mechanisms. The former produce a sudden lost of the cold and hot coronal gas of a galaxy. In a few Myr the gas of the galaxy could be swept by the ram-pressure mechanism. This would produce a sudden stop of the star formation and the formation of a PSB galaxy. In contrast, the quenching time-scale for the starvation mechanism is much larger. Several Gyr are needed in order to stop the star formation and produce a PSB galaxy by starvation. This indicate that if starvation is the main mechanism producing the PSB phase in A\ 85 galaxies, they should be mainly located in the external regions of the cluster since this phase is short (less than 1 Gyr) and galaxies expend more time around the apocenter of their orbits. This is contrary  to what is observed for the PSB galaxies of A\ 85. About 50$\%$ of the PSB galaxies in A \ 85 are located within 0.5$r_{200}$. They are even more centrally concentrated than PAS galaxies. Indeed, those PSB with the strongest post-starburst features are the closest to the cluster centre. These sPSB galaxies are those that are caught in earlier stages of their quenching. This means that they have stopped their star formation in the last $\sim 0.5$ Gyr, indicating that the quench of the star formation in the PSB galaxies has occurred close to the centre of the cluster when the galaxies are passing by the pericenter of their orbits. Galaxies passing through the pericenter of the orbits have the largest relative velocities and they are embedded in the densest gas regions of the cluster. In these regions closer to the cluster centre is where the ram-pressure is strongest. This result points towards the ram-pressure stripping as the responsible of the formation of the PSB galaxies in A\ 85. 

Following \cite{jaffe2015} we have analyzed the ram-pressure stripping in A\ 85. This was done by creating a simple model based on the ram-pressure description given by \cite{gunn1972}. Thus, the intensity ($\eta$) of the ram-pressure  exerted by the intra-cluster medium on an infalling galaxy is given by the ratio $\eta = P_{ram}/\Pi_{gal}$, where $P_{ram}$ and $\Pi_{gal}$ are the ram-pressure and the anchoring self-gravity provided by the galaxy, respectively \citep[see][]{hernandez2014, jaffe2015}. The ram-pressure is defined as: $P_{ram} = \rho_{ICM} v_{gal}^{2}$, where $\rho_{ICM}$ is the intra-cluster gas density, and $v_{gal}^{2}$ is the velocity of the galaxy. In the case of A\ 85 we obtained $\rho_{ICM}$ from \cite{durret2005}. The anchoring pressure is given by: $\Pi_{gal}= 2 \pi G \Sigma_{s} \Sigma_{g}$, where $\Sigma_{s}$ and $\Sigma_{g}$ are the density profiles of the stellar and gaseous disks of the galaxies. Exponential laws have been assumed for these profiles. We have predicted the region of the phase-space diagram where ram-pressure is expected to strip the gas off the galaxies with $10^{8}, 10^{9}$ and $10^{10}$ $M_{\odot}$ (see Fig. \ref{velpsb}). These regions were obtained assuming scale lengths ($R_{d}$) for the stellar disks given by: 0.8, 0.9, and 1.7 kpc for galaxies with 10$^{8}, 10^{9}$ and $10^{10}$ M$_{\odot}$, respectively. These scales lengths were obtained from to the mean values of the effective radius ($R_{e}$) measured for the galaxies in A\ 85 in the three mass bins, and assuming $R_{e} = 1.67 R_{d}$. We have also considered the same relations as \cite{jaffe2015} between the mass and scale lengths of the gas and stars in the galaxies. According to this model the vast majority of the cluster member galaxies with $10^{8}$ M$_{\odot}$ would be stripped by ram-pressure. In contrast, the stripped region of galaxies with $10^{10}$ M$_{\odot}$ is very small. Our sample of PSB galaxies in A\ 85 turned to have stellar masses smaller than $6.05 \times 10^{9}$ M$_{\odot}$, being $75\%$ of them with masses smaller than $10^{9}$ M$_{\odot}$. These masses and the position of the galaxies in the phase-space diagram point towards ram-pressure stripping is the mechanism producing the PSB phase of the cluster members of A\ 85.

One caveat to this result is the number of EL galaxies that are located in the infall region of the phase-space diagram of A\ 85. These galaxies with masses between $10^{9} - 10^{8}$ M$_{\odot}$ should be stripped according with the model discussed before. This could be explained by the simplicity of the ram-pressure model considered before. The efficiency of  ram-pressure is more complex depending on other parameters like  the galaxy orbit or the galaxy inclination during the infall procces  \citep[see, ][]{vollmer2001}. The difference in the orbits between blue and red galaxies in A\ 85 has been pointed by \cite{aguerri2017}. In particular, they showed that red galaxies in A\ 85 are located in more radial orbits than the blue ones. This difference in the orbital properties is  more pronounced for dwarf galaxies. Tangencial orbits are expected to produce less damage than radial ones unless their radii are very small. This means that EL galaxies would survive in the infall cluster region if they are located in circular orbits, no strongly affected by the ram-pressure.

\subsection{Pre-processing in the infall cluster region around A\ 85}

Pre-processing transform galaxies living in small groups prior to the cluster infall \citep[][]{fujita2004}. It has been invoked previously in the literature in order to explain some galaxy properties such as the fraction of passive galaxies at large clustercentric distances \citep[e.g.,][]{haines2015},  the prevalence of S0 galaxies in large clusters \citep[][]{kodama2001,wilman2009} or the dependence of the satellite star formation rate and morphology with the group dynamics \citep[][]{roberts2017}. 

The PSB galaxy population trace the sample of recently quenched galaxies. We found that for the full sample of 711 galaxies with $\rm{V} < 32000$ km s$^{-1}$ about 4.5$\%$ of this populations are PSB galaxies. These recently quenched galaxies are located both inside the A\ 85 cluster (80$\%$) and in the non-cluster member population (20 $\%$). These numbers show that although the bulk of the quenching in galaxies is taking place inside A\ 85, there is also some quenching episodes outside the cluster environment. These quenched galaxies out from A\ 85 indicates the presence of pre-processing in this galaxy population. 

In addition to this, five of the PSBs in A\ 85 are asociated with the galaxy substructures obtained by  \citep[][]{yu2016}. One of them is located in the infall region and the remaining four in the virialized one. This means that for these cases some pre-processing of the galaxies in the substructures could quench their star formation and produce the PSB phase. 

This points towards that a fraction of the PAS galaxies observed today inside A\ 85 have probably been pre-processed in small groups previous or in the process of infall into the cluster.

\section{Conclusions}

We have analyzed the spectral properties of 711 galaxies with $\rm{V} < 32000$ km s$^{-1}$ in the direction of A\ 85. The main results are:

\begin{itemize}
\item The large-scale galaxy distribution around A\ 85 is a complex structure formed by the cluster itself and several galaxy groups (A\ 89b and A\ 89c) connected by a filamentary structure.
\item There is a critical galaxy density indicating a colour transformation with the environment. Galaxies located at $\Sigma < \Sigma_{c}$ ($\Sigma_{c} \sim 30$ Gal Mpc$^{-2}$) show bluer $g-r$ colours. In contrast, galaxies at higher densities do not show a colour variation with the density.
\item The colour - density relation is only observed for galaxies fainter than $M_{r} = -20.0$. This indicates that the environment plays an important role in the quenching of faint galaxies. 

\item The galaxies were classified in three groups according to their EWs in [OII] and H$\delta$. Galaxies with no emission in [OII] (EW([OII]) $> -5$ \AA) and EW(H$\delta) < 5$ \AA\ are considered as PAS ones. EL galaxies are those with emission in [OII] (EW([OII]) $< -5$ \AA). PSB galaxies are those with EW([OII])$ > -5$ \AA\ and EW(H$\delta) > 5$ \AA.
\item For the full sample of galaxies we have 68.5 $\%$, 27.0$\%$, and 4.5$\%$ of PAS, EL and PSB galaxies, respectively. The percentages for the non-cluster member population are: 41.0$\%$, 56.0$\%$, and 3.0$\%$ for PAS, EL, and PSB galaxies. For the cluster members we have: 78.8$\%$, 16.0$\%$, and 5.2$\%$ of PAS, EL, and PSB galaxies.
\item Most of the PSB galaxies ($80\%$ for the full and cluster member samples, and 70$\%$ for the non-member sample) are dwarf ($M_{r} > -18.0$ or M$_{*} < 10^{9}$ M$_{\odot}$) objects. In addition, no PSB galaxy was found with $M_{r} < -20.0$ or more massive than $6.05 \times 10^{9}$ M$_{\odot}$. For the cluster member sample, PSB galaxies are the faintest galaxy population. EL and PSB galaxies have similar luminosities in the non-cluster member sample.
\item Independently of the environment, PSB galaxies show  $g - r$ stellar colours intermediate between PAS and EL ones. 
\item PSB galaxies on the cluster member sample are located in high-density environments similar to PAS galaxies. For the non-cluster member sample EL and PSB galaxies share the same galaxy density environment. In all samples PAS galaxies are located in denser environments than EL ones.
\item EL galaxies in A\ 85 show larger values of $|V-V_{c}|/\sigma_{c}$ than PAS ones. In addition, PSB and sPSB galaxies show similar values of $|V-V_{c}|/\sigma_{c}$ than PAS ones.
\item PAS galaxies are located closer to the cluster centre than EL ones. In contrast, sPSB and PSB galaxies are located closer than PAS galaxies. There is also a relation between the value of the EW(H$\delta$) and the location within the clusters. Thus, the galaxies with the largest values of EW(H$\delta$) (sPSB galaxies) are located closer to the cluster centre.
\item PSB galaxies show a disc-like structure independent of the environment where they live.
\end{itemize}

These results suggest that the mechanisms driving the formation of PSB galaxies depends on the environment. In low-density environments, such as filaments or small galaxy groups, EL galaxies  could suffer gas-rich minor mergers or accretions that activate a massive starburst and consume most of their gas reservoir, then producing a PSB galaxy. For high density environments, like A\ 85 the PSB galaxies are formed when EL galaxies in radial orbits, similar to PAS ones, pass near to the pericenter of their orbit. Then, the ram-pressure removes  their gas content and suddenly stop their star formation producing a PSB galaxy. 

This study has been done in the unique environment of A\ 85. Nevertheless, a similar detailed study  in a larger sample of individual nearby galaxy clusters remains to be done. We plan to obtain deep spectroscopy in a sample of several galaxy clusters covering a large range of masses and dynamical states. This will allow us to study the variation of the quenching in different cluster environments.

\section*{acknowledgements}
   We would like to thank to Dr. Claudio Dalla Vecchia for useful comments during the writing process of this manuscript. JALA, JMA, and IA want to thank the support of this work by the Spanish Ministerio de Economia y Competitividad (MINECO) under the grant AYA2013-43188-P. This research has made use of the Thirteen Data Release of SDSS, and of the NASA/IPAC Extragalactic Database which is operated by the Jet Propulsion Laboratory, California Institute of Technology, under contract with the National Aeronautics and Space Administration. The WHT and its service programme are operated on the island of La Palma by the Isaac Newton Group in the Spanish Observatorio del Roque de los Muchachos of the Instituto de Astrof\'{\i}sica de Canarias.







\bsp	
\label{lastpage}
\end{document}